\documentclass[preprint,showkeys,secnumarabic,amsfonts,showpacs,amsmath,amssymb]{revtex4}
\usepackage[dvips]{color}
\usepackage{epsfig}
\usepackage{amsmath}
\usepackage{graphicx}

\begin{document}

\title{Stability analysis of agegraphic dark energy in Brans-Dicke cosmology}

\author{H. Farajollahi$^{1,2}$}\email{hosseinf@guilan.ac.ir}  \author{J. Sadeghi$^{3,4}$}\author{M. Pourali$^{1}$} \author{A. Salehi$^{1}$}
\affiliation{Department of Physics, University of Guilan, Rasht, Iran}
\affiliation{School of Physics, University of New South Wales, Sydney, NSW, 2052, Australia}
\affiliation{Sciences Faculty, Department of Physics, Mazandaran University, P .O .Box 47416-95447, Babolsar, Iran}
\affiliation{Institute for Studies in Theoretical Physics and Mathematics (IPM), P.O.Box 19395-5531, Tehran, Iran}

 \begin{abstract}
Stability analysis of agegraphic dark energy in Brans-Dicke theory is presented in this paper. We constrain the model parameters with the observational data and thus the results become broadly consistent with those expected from experiment. Stability analysis of the model without best fitting shows that universe may begin from an unstable state passing a saddle point and finally become stable in future. However, with the best fitted model, There is no saddle intermediate state. The agegraphic dark energy in the model by itself exhibits a phantom behavior. However, contribution of cold dark matter on the effective energy density modifies the state of teh universe from phantom phase to quintessence one. The statefinder diagnosis also indicates that the universe leaves an unstable state in the past, passes the LCDM state and finally approaches the sable state in future.

\end{abstract}

\maketitle

\section{Introduction\label{Int}}

Cosmological probes such as Cosmic Microwave Background (CMB) \cite{Dunkley}, \cite{Komatsu}, Supernova type Ia (SNIa)\cite{Knop}, \cite{Riess}, Weak Lensing\cite{Leauthand}, Baryon Acoustic Oscillations (BAO) \cite{Parkinson}, 2dF Galaxy Redshift Survey (2dFGRS) \cite{Cole} at low redshift and
DEEP2 redshift survey \cite{Yan} at high redshift, have given us cross-checked data to determine
cosmological parameters with high precision. These parameters indicate that our universe is nearly spatially flat, homogeneous and isotropic at large
scale, i.e. a Friedmann-Robertson-Walker (FRW) with zero curvature, and has entered an accelerating epoch at about $z \approx 0.46$ \cite{Riess}. In addition, from $\Lambda CDM$ model, the universe consists of $0.046$ baryonic matter,
$0.228$ non-relativistic unknown matter, called as cold dark matter (CDM), and a significant amount
of $0.726$ smoothly distributed dominant dark energy (DE) \cite{Komatsu}.

The equation of state (EoS) of DE, is the main parameter which determines
the gravitational effect on the dynamics of the universe, and can be measured from
observations without need to have a definite model of DE. Strong evidences suggest that the
EoS of DE lies in a narrow range around $w \approx -1$ with a smooth behavior \cite{Riess}, \cite{Amanullah}. We can classify the EoS parameter of
DE with respect to the barrier $w = -1$, namely the phantom divide line (PDL )\cite{Cai}. That is,
DE with the EoS parameter of $w = -1$ is named for the cosmological constant, $\Lambda$, with a constant energy
density. The case with dynamical EoS parameter of $w\geq -1$ and $w \leq -1$ are respectively referred to quintessence and phantom cosmology \cite{Gonzalez} \cite{Carroll} \cite{Caldwell}.

A fundamentally important problem in modern cosmology is the current state of the universe to be in phantom or quintessence phase? A transition from phantom to quintessence phase and reverse plus variation of  EoS parameter being close to -1 imply that universe may evolve into a finite--time future singularity in future \cite{Nojiri4} \cite{Elizalde3}\cite{Capozziello2}. Different cosmological models predict different scenarios based on their own assumptions and criterions \cite{Nojiri5}\cite{Nojiri6}\cite{Nojiri7}. Among them, the inclusion of CDM in the formalism, and its coupling to DE and or geometry to explain phantom transition, varying EoS parameter and future singularity are favored by many authors \cite{Nojiri3}\cite{Ito}.

Although, the issue of DE, its energy density and EoS parameter is still an unsolved problem in classical gravity, it may be in the context of quantum gravity that we achieve a more inclusive insight to its properties \cite{Cohen}. The holographic dark energy (HDE) model is an alternative to formulate DE within the framework of quantum gravity \cite{Hsu},\cite{Li}. The holographic principle expresses that the number of degrees of freedom
describing the physics inside a volume (including gravity) is confined by the
area of the boundary which encloses the volume and therefore related to the entropy scales with the boundary \cite{Hooft}. The entropy scales like the area rather than the volume, and thus the fundamental degrees of freedom rendering the system are characterized by an effective quantum field theory in a box of size $L$ with one fewer space dimensions and with planck-scale UV cut-off $\Lambda$ \cite{Hooft}.

On the other hand, a viable alternative approach to dark energy that naturally gives rise to late time accelerating solutions as well as early time inflation is extended modified gravity (For a review see ). Among extended theories of gravity, scalar tensor theories are the best motivated and most promising candidates to general relativity. They are compatible with the observational findings without the need of additional components and may resolve the coincidence problem. The birth of scalar tensor theories is marked by the work of Jordan in which a scalar field coupled to both curvature and matter Lagrangian. Later, Brans-Dick demanded the decoupling of matter Lagrangian from scalar field to guarantee the validity of the weak equivalence principle. Though, the "decoupling of the scalar" (called as Brans-Dicke problem) is the most serious shortcoming of this approach, it has been argued that the cosmic acceleration can be generated by either assuming a time dependent Brans-Dicke parameter $\omega{t}$, or adding a potential term to the lagrangian \cite{Elizalde}\cite{Nojiri1}\cite{Nojiri2}\cite{Capozziello}.

Alternative, with regards to HDE, the scalar-tensor theories have been widely used to explain the late time acceleration of the universe \cite{Sahoo}--\cite{farajollahi}. In particular, in Brans-Dicke (BD) theory \cite{Brans}\cite{Mathiazhagan}\cite{La}\cite{Steinhardt}\cite{Sahoo}, where the gravitational constant varying as inverse
of a time dependent scalar field, cosmic acceleration is predicted.

While the successful HDE model explains the observational data, more recently, a new dark energy model, dubbed "agegraphic
dark energy" (ADE) model, has been proposed by Cai \cite{R}. The ADE is also related to the holographic
principle of quantum gravity and considers the uncertainty relation of quantum mechanics together with the gravitational effect in general relativity.

In this work we present the ADE in Brans-Dicke theory after performing stability analysis and constraining the model parameters with observational data. As noted by \cite{Banerjee}, the acceptable cosmic acceleration in BD theory can be realised only if the BD parameter is time dependent, or a potential term included in the action. However, in this work, we reproduce current universe acceleration in the context of BD theory without adding potential function or assuming time dependent BD parameter. Instead, with the agegraphic description of dark energy, the best fitted effective EoS parameter exhibits current cosmic acceleration in quintessence phase and matter dominated phase in the far past. Together with statefinder parameters \cite{Sahni} \cite{Alam}-\cite{H. Farajollahi}\cite{Linder}, we also discuss the dynamics of the model in different cosmological epochs.


\section{The model \label{FE}}

The Brans-Dicke action is given by,
\begin{eqnarray}
 S=\int{
d^{4}x\sqrt{g}\left(\frac{\phi ^2}{8\omega}
{R}-\frac{1}{2}g^{\mu\nu}\nabla_{\mu}\phi \nabla_{\nu}\phi
+L_M \right)},\label{act1}
\end{eqnarray}
where $\omega$ is a dimensionless coupling constant which determines the coupling between gravity
and BD scaler field, ${R}$ is Ricci scaler, $\phi(x^\mu)$ is the BD scalar
 field and $L_{M}$ Lagrangian of matter field.  We assume the metric is in the form of,
\begin{eqnarray}
 ds^2=-dt^2+a^2(t,y)\left(\frac{dr^2}{1-kr^2}+r^2d\Omega^2\right),\label{metric}
 \end{eqnarray}
where space time is homogeneous and isotropic (FRW  universe). In the metric, $k$ is curvature parameter with $k = -1, 0, 1$ corresponding to open, flat, and
closed universes, respectively. The BD scaler field $\phi$ and the scale factor $a$, are functions of $t$ only. Variation of action (\ref{act1}) with respect to metric (\ref{metric}) yields the following field equation,
\begin{eqnarray}
\frac{3}{4\omega}\phi^2\Big(H^2+\frac{k}{a^2}\Big)-\frac{1}{2}\dot{\phi}^2+\frac{3}{2\omega}H\dot{\phi\phi}=\rho_m+\rho_\Lambda,\label{BD1}
\end{eqnarray}
\begin{eqnarray}
\frac{-1}{4\omega}\phi^2\Big(2\frac{\ddot{a}}{a}+H^2+\frac{k}{a^2}\Big)- \frac{1}{\omega}H\dot{\phi\phi}-\frac{1}{2\omega}\ddot{\phi}\phi-\nonumber\\ \frac{1}{2}\Big(1+\frac{1}{\omega}\Big)\dot{\phi}^2=p_\Lambda,\label{BD2}
\end{eqnarray}
\begin{eqnarray}
\ddot{\phi}+3H\dot{\phi}-\frac{3}{2\omega}\Big(\frac{\ddot{a}}{a}+
H^2+\frac{k}{a^2}\Big)\phi=0,\label{BD3}
\end{eqnarray}
where $H=\frac{\dot{a}}{a}$.
Moreover, the conservation equations for dark energy and matter field in the universe are respectively,
\begin{eqnarray}
&&\dot{\rho}_\Lambda+3H\rho_\Lambda(1+w_\Lambda)=0,\label{consq}\\
&&\dot{\rho}_m+3H\rho_m=0, \label{consm}
\end{eqnarray}

Next, we apply the ADE model in Brans-Dicke theory. The ADE model with dark energy density is given by
\begin{eqnarray}
\rho_{\Lambda}=\frac{3{n}^2M_P^2}{T^2},\label{NADE}
\end{eqnarray}
where $T$ is age of the universe and given by
\begin{eqnarray}
T=\int{\rm d}t=\int_0^a\frac{{\rm
d}a}{Ha}.\label{eta}
\end{eqnarray}
In the framework of Brans-Dicke cosmology, we write the ADE of quantum fluctuations in the
universe as
\begin{eqnarray}\label{rho1n}
\rho_{\Lambda}= \frac{3n^2\phi^2 }{4\omega T^2}.
\end{eqnarray}
In the next section we constrain the model with the data from observational probe, and study the stability of the dynamical system.

\section{Stability analysis and observational constraints}

The structure of the dynamical system can be studied via phase plane analysis. By introducing
the following dimensionless variables,
\begin{eqnarray}\label{HL}
\Omega_m=\frac{4\omega\rho_m}{3\phi^2
H^2}, \ \ \ \Omega_k=\frac{k}{H^2 a^2}, \ \ \
\Omega_\Lambda=\frac{n^2}{H^2T^2}.
\end{eqnarray}
We assume that the BD scalar field $\phi$ is in power law of the scale factor $a(t)$, i.e. $\phi(t)\propto a^\alpha$. Therefore, the field equations in terms of new dynamical variables become,
\begin{eqnarray}
{\Omega'_{m}}&=&-\Omega_{m}(3+2\alpha+2\frac{\dot{H}}{H^2})\label{fs1}\\
\Omega'_{\Lambda}&=&-2\Omega_{D}(\frac{\sqrt{\Omega_{D}}}{n} +\frac{\dot{H}}{H^2})\label{fs2}\\
\Omega'_{k}&=&-\Omega_{k}(1+2\frac{\dot{H}}{H^2})\label{fs3}
 \end{eqnarray}
 where
\begin{eqnarray}\label{hd}
\frac{\dot{H}}{H^2}=-\frac{3}{2(3\alpha+3)}\Big[(2\alpha+1)^2+2\alpha(\alpha\omega-\frac{3}{2})
\nonumber\\
+\Omega_k-1+3\Omega_{\Lambda}(-1-\frac{2\alpha}{3}+ \frac{2\sqrt{\Omega_{\Lambda}}}{3n})\Big]-\frac{3}{2}
 \end{eqnarray}
By using the Friedman constraint in terms of the new
dynamical variables:
\begin{eqnarray}
\Omega_{k}=\Omega_{m}+\Omega_{\Lambda}-2\alpha(1-\frac{\alpha\omega}{3})-1
\end{eqnarray}
the equations (\ref{fs1}-\ref{fs3}) reduce to,
\begin{eqnarray}
\Omega'_{m}&=&\Omega_{m}\Big(-2\alpha+\frac{3}{(3\alpha+3)}\Big((2\alpha+1)^2+\Omega_{m} \nonumber\\
&+&2\alpha(\alpha\omega-\frac{3}{2})+\Omega_{\Lambda}
-2\alpha(1-\frac{\alpha\omega}{3})
 \nonumber\\ &-&2
+3\Omega_{\Lambda}(-1-\frac{2\alpha}{3}
+\frac{2\sqrt{\Omega_{\Lambda}}}{3n})\Big)\Big)
\end{eqnarray}
\begin{eqnarray}
\Omega'_{\Lambda}&=&\Omega_{\Lambda}\Big[\frac{-2\sqrt{\Omega_{\Lambda}}}{n}+\frac{3}{(3\alpha+3)}
\Big((2\alpha+1)^2\nonumber\\
&+&2\alpha(\alpha\omega-\frac{3}{2})-2\alpha(1-\frac{\alpha\omega}{3})
-2 +\Omega_{m}+\Omega_{\Lambda}
\nonumber\\ &+&3\Omega_{\Lambda}(-1-\frac{2\alpha}{3}  +\frac{2\sqrt{\Omega_{\Lambda}}}{3n})\Big)+3\Big]
\end{eqnarray}
where prime means derivative with respect to $ln(a)$. In stability formalism, by simultaneously solving $\Omega'_{m}=0$ and $\Omega'_{\Lambda}=0$, we find the fixed points (critical points). However, in this work we first constrain the model with observational data for distance modulus using $\chi^2$ method. The model parameters we best fit are $\alpha$, and the ADE, BD and Hubble parameters, $n$, $\omega$ and $h_0$ respectively. The observational data we use is the most recent SNe Ia data given by $557$ data points. The $chi^2$ method will be employed to best fit the model parameters with the observational data. Figs 1 and 2 show the best fitted 2-dim likelihood and confidence level for the parameters $\alpha$, $h_0$, $n$ and $\omega$.

\begin{tabular*}{.5 cm}{cc}
\includegraphics[scale=.3]{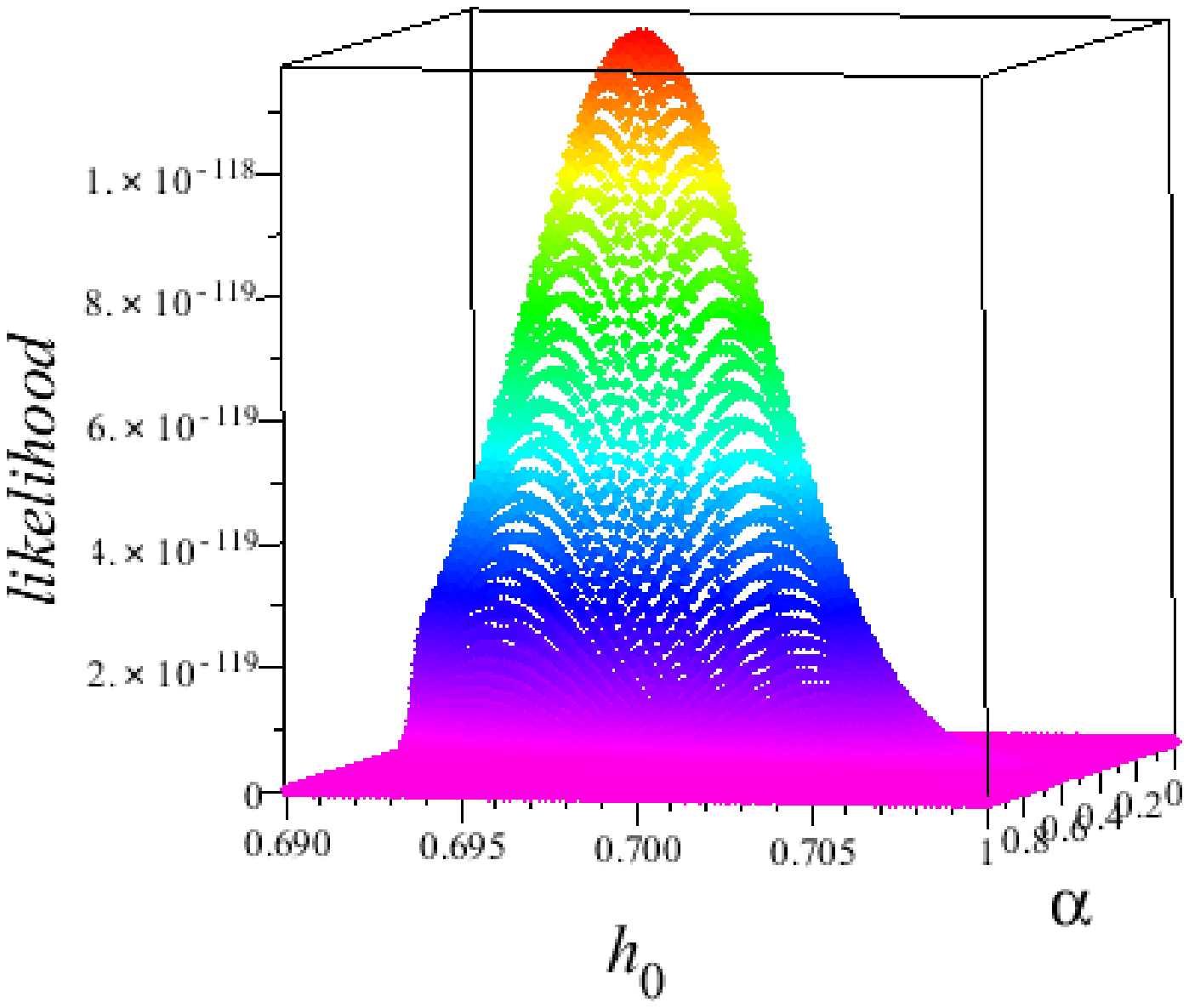}\hspace{0.1 cm} \includegraphics[scale=.2]{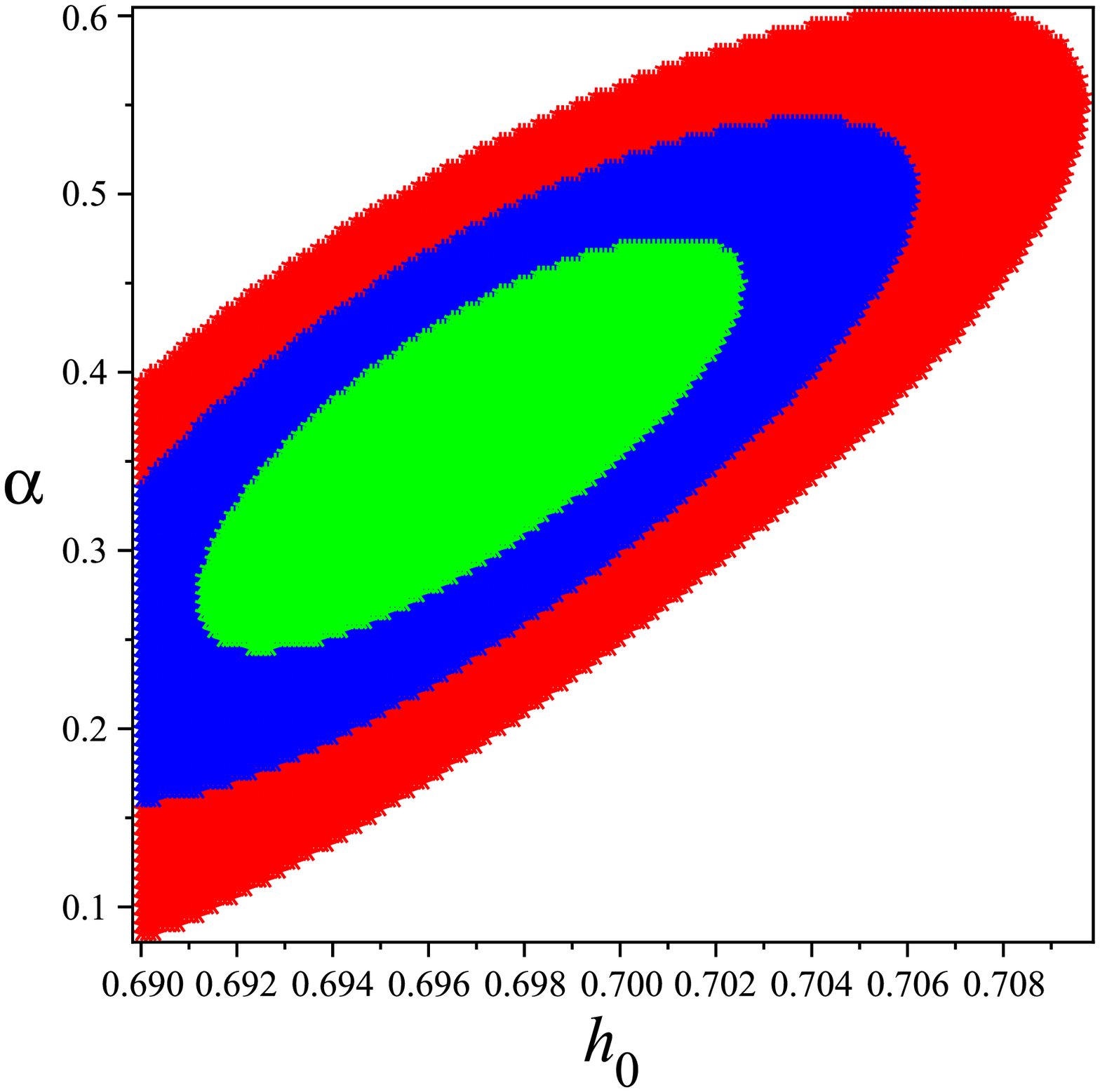}\hspace{0.1 cm}\\
Fig. 1:The best-fitted 2-dim likelihood and confidence level for $\alpha$ and $h_0$\\
\end{tabular*}\\
\begin{tabular*}{.5 cm}{cc}
\includegraphics[scale=.3]{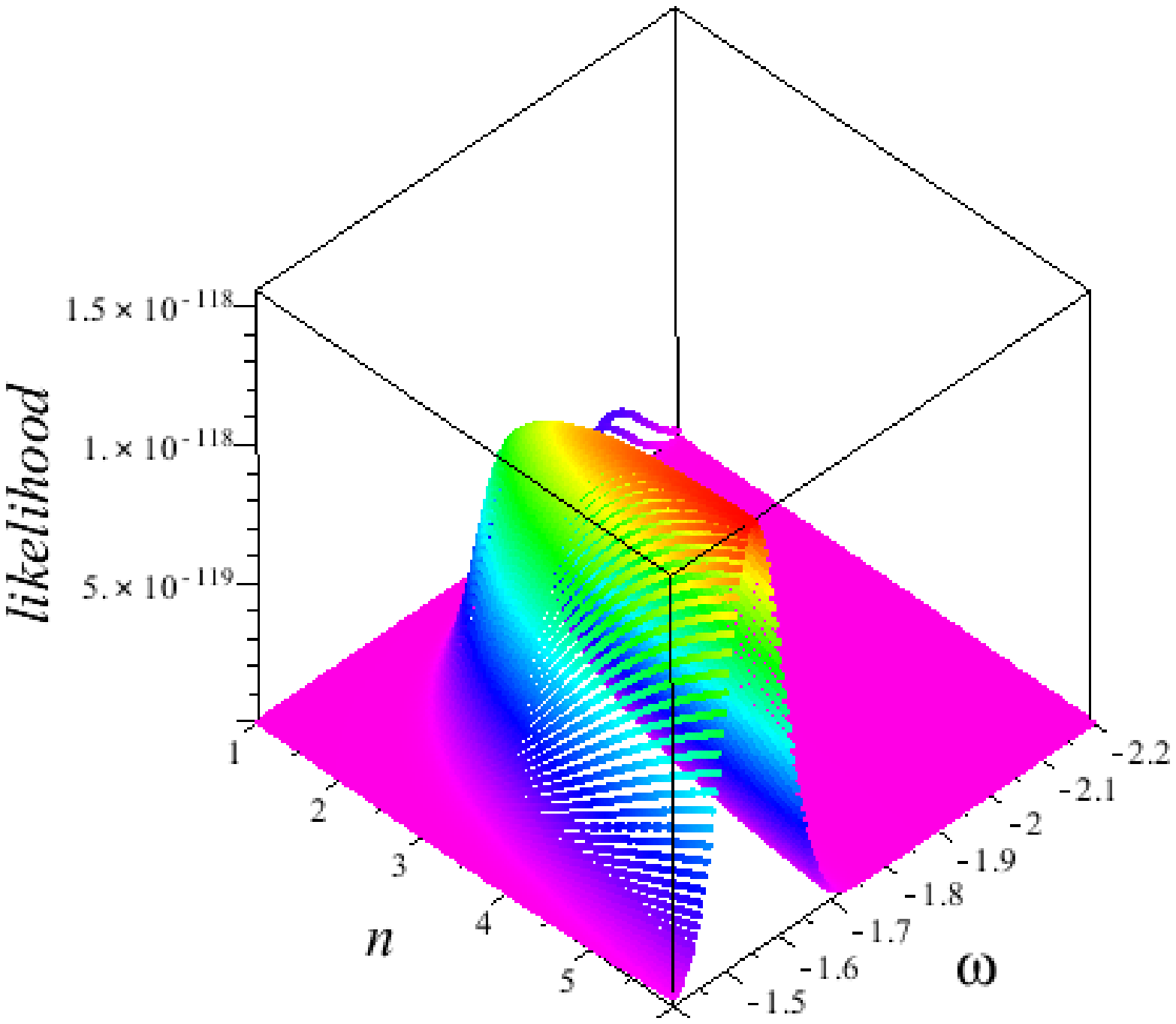}\hspace{0.1 cm} \includegraphics[scale=.2]{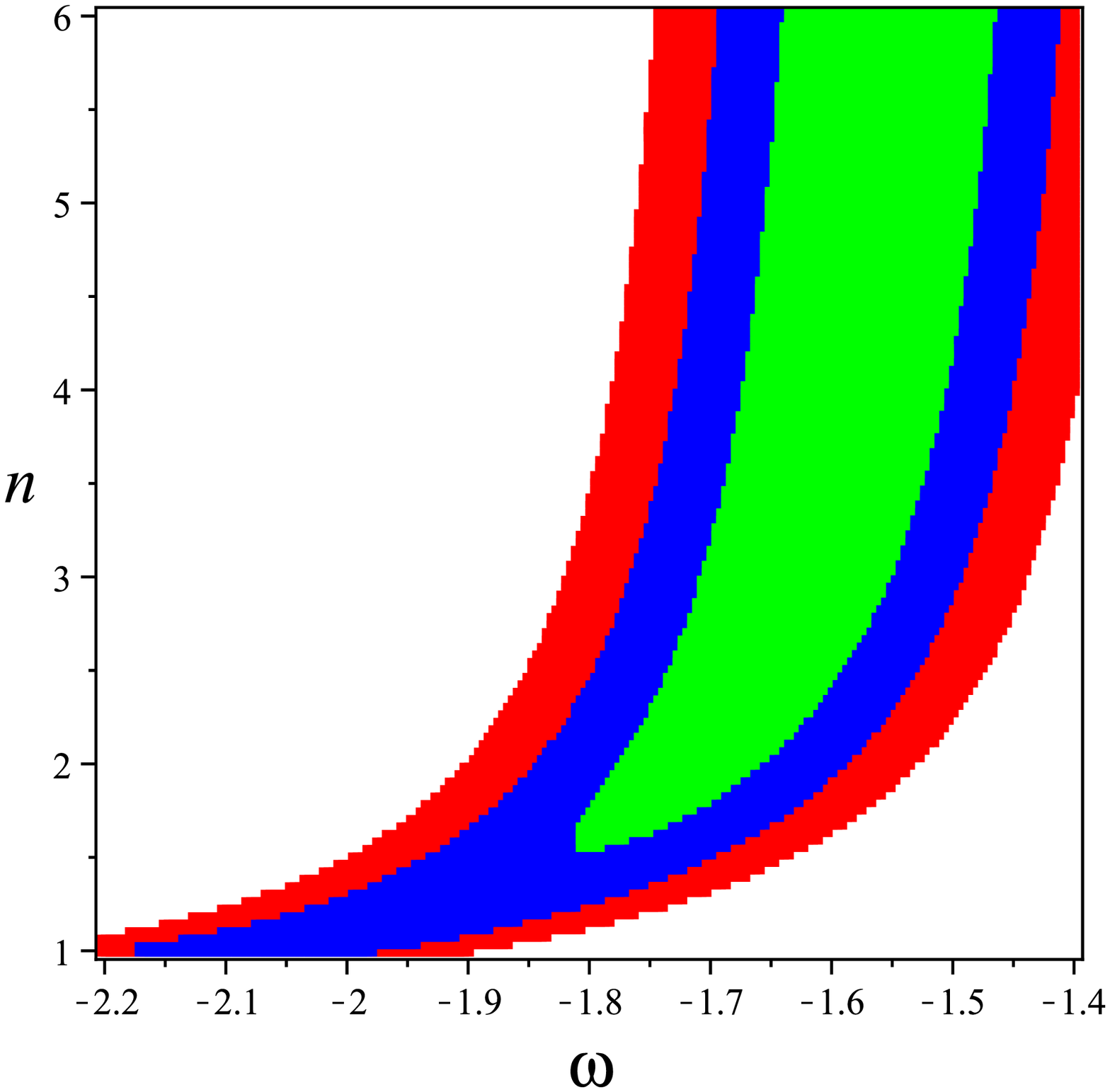}\hspace{0.1 cm}\\
Fig. 2:The best-fitted 2-dim likelihood and confidence level for $n$ and $\omega$\\
\end{tabular*}\\

We the values of the best fitted parameters we study the stability of the model. We solve the dynamical system and obtain the critical points. From eigenvalues, the stability of the critical points are given in Table I.
\begin{table}[ht]
\caption{Best fitted critical points} 
\centering 
\begin{tabular}{c c c c c c c c c} 
\hline\hline 
point  &  $(\Omega_{m},\Omega_{\Lambda})$ \ & eigenvalue\ &stability \\ [4ex] 
\hline 
$P_{1}$ & \ (4.7,0) & (2.8,4.4) & unstable   \\
\hline 
$P_{2}$ & \ (0,0.6) \ & (-1.1,-3.9) & stable    \\
\hline 
$P_{3}$ & \ (0,0) \ & (-2.8,1.6) & saddle   \\
\hline 
$P_{4}$ & \ (0,27.8) \ & (1.9,-0.9) & saddle    \\
\hline 
$P_{5}$ & \ (-38.8,43.6) \ & $(3\times10^-7,2.8)$ & unstable   \\
\hline 
\end{tabular}
\label{table:1} 
\end{table}\
From Table I, there are five critical points where only one of them is stable. In the 2-dim phase space ( Fig 3), the best fitted trajectory leaves unstable critical point $P_1$, towards the stable critical point $P_2$ is shown. Note that the critical points are presenting the state of the universe. By small perturbation of the dynamical variables, the green color best fitted trajectory in the phase plane illustrates that the dynamical universe begins from an unstable state $P_1$ in the past in the phase plane and approaches the stable state $P_2$ in future.

\begin{tabular*}{.5 cm}{cc}
\includegraphics[scale=.3]{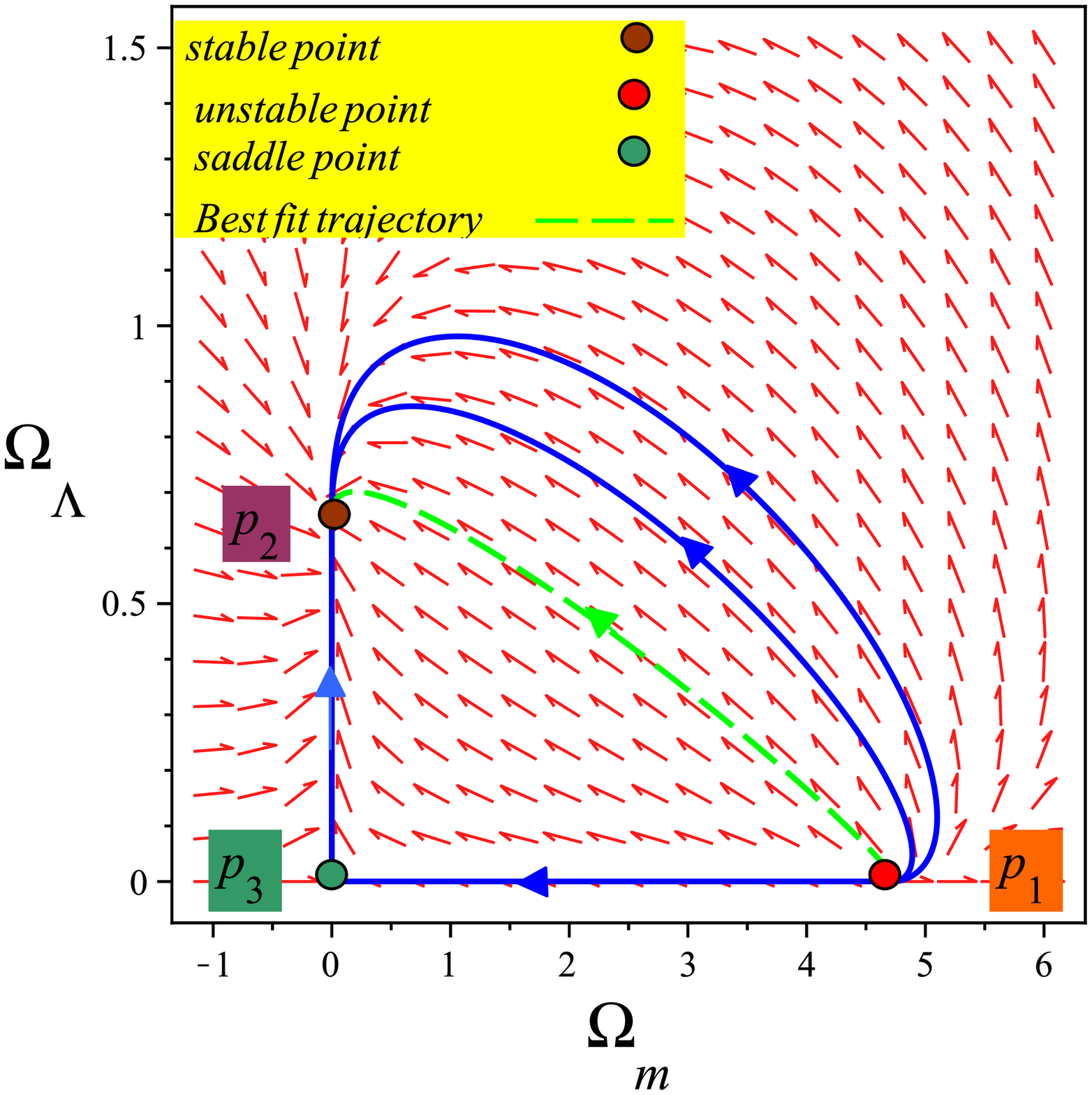}\hspace{0.1 cm}\\
Fig. 3:The 2-dim phase plane corresponding to  \\ the critical point\\
\end{tabular*}\\

At this point, from the graph of phase plane not enough physical information can be found. Our original model is presented in term of new dimensionless variables with not enough physical justification. In addition, the phase space shows the behavior of the system in terms of these new variables. One needs to represent the physical quantities in terms of these variables or alternatively switch back into the original variables to understand the cosmological consequences. More specifically, in the following we discuss the cosmological parameters such as EoS parameter and statefinders that are among the most practical parameters to discriminate cosmological models. The effective EoS parameter, $w_{eff}$, agegraphic EoS parameter, $w_\Lambda$, are in particular given in terms of new dimensionless variables as
\begin{eqnarray}\label{weff}
w_{eff}=\frac{1}{3\alpha+3}\Big((2\alpha+1)^2+2\alpha(\alpha\omega-\frac{3}{2})+\Omega_m+\Omega_\Lambda\nonumber\\
-2\alpha(1-\frac{\alpha\omega}{3})-2+3\Omega_\Lambda(-1-\frac{2\alpha}{3}+\frac{2\sqrt{\Omega_\Lambda}}{3n})\Big)
\end{eqnarray}
\begin{eqnarray}\label{wl}
w_\Lambda=-1-\frac{2\alpha}{3}+\frac{2\sqrt{\Omega_\Lambda}}{3n}
\end{eqnarray}
Also, the statefinder parameters in terms of deceleration parameter $q$
and $\frac{\ddot{H}}{H^2}$ are defined as $r = \frac{\ddot{H}}{H^3}-3q-2$
and $s =\frac{(r-1)}{3(q-\frac{1}{2})}$.

In Table II, the values of the effective EoS and deceleration parameters in addition to statefinder parameters $\{r, s\}$ are shown at the critical points and also at current epoch.

\begin{table}[ht]
\caption{The values of cosmological parameters} 
\centering 
\begin{tabular}{c c c c c c c c c} 
\hline\hline 
parameter  &  $w_{eff}$ \ & $q$ &  $r$ \ & $s$ \\ [4ex] 
\hline 
current value & \ $-0.8$ & $-0.695$ & \ $0.3$ & $0.17$ \\
\hline 
unstable Point & \ $0.467$ & $1.2$ & \ $4.08$ & $-1.34$ \\
\hline 
stable Point & \ $-0.82$ & $-0.732$ & \ $0.35$ & $0.14$ \\
\hline 
\end{tabular}
\label{table:1} 
\end{table}

The effective EoS parameters is shown in Fig. 4)top. The parameter begins to change from unstable critical point in the past moving towards the stable critical point in the future. From the graph, its current value is $-0.8$. A comparison of the effective EoS parameter and ADE EoS parameter is given in Fig. 4)below. From the graph we see that $w_\Lambda$ is always less than -1 and thus represents the phantom regime. The contribution of CDM in the universe causes the effective EoS parameter levels up from phantom phase to quintessence phase. Note that these trajectories are for the best fitted parameters with the observational data and therefore are physically significant.

\begin{tabular*}{2.5 cm}{cc}
\includegraphics[scale=.3]{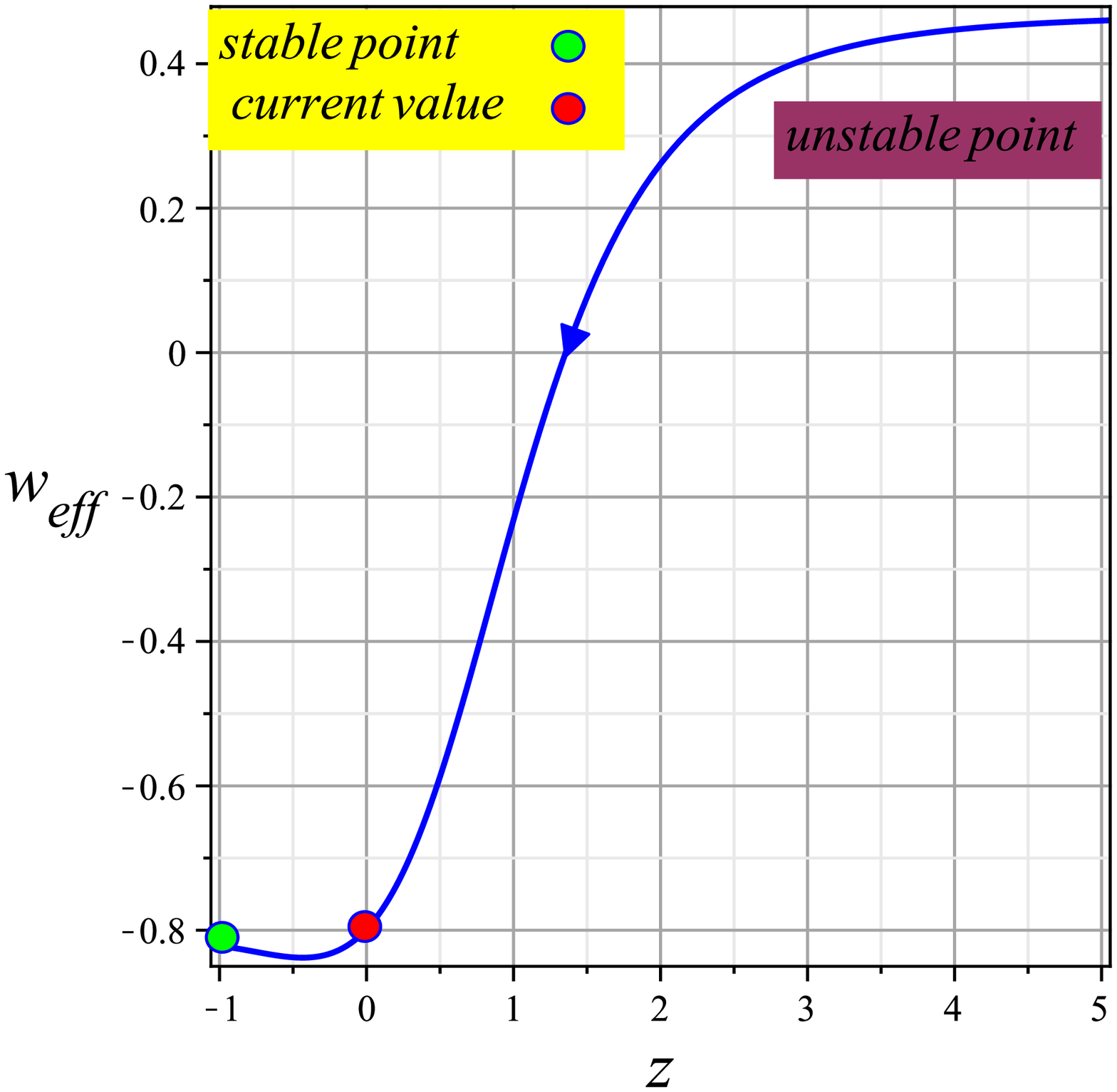}\hspace{0.1 cm} \includegraphics[scale=.3]{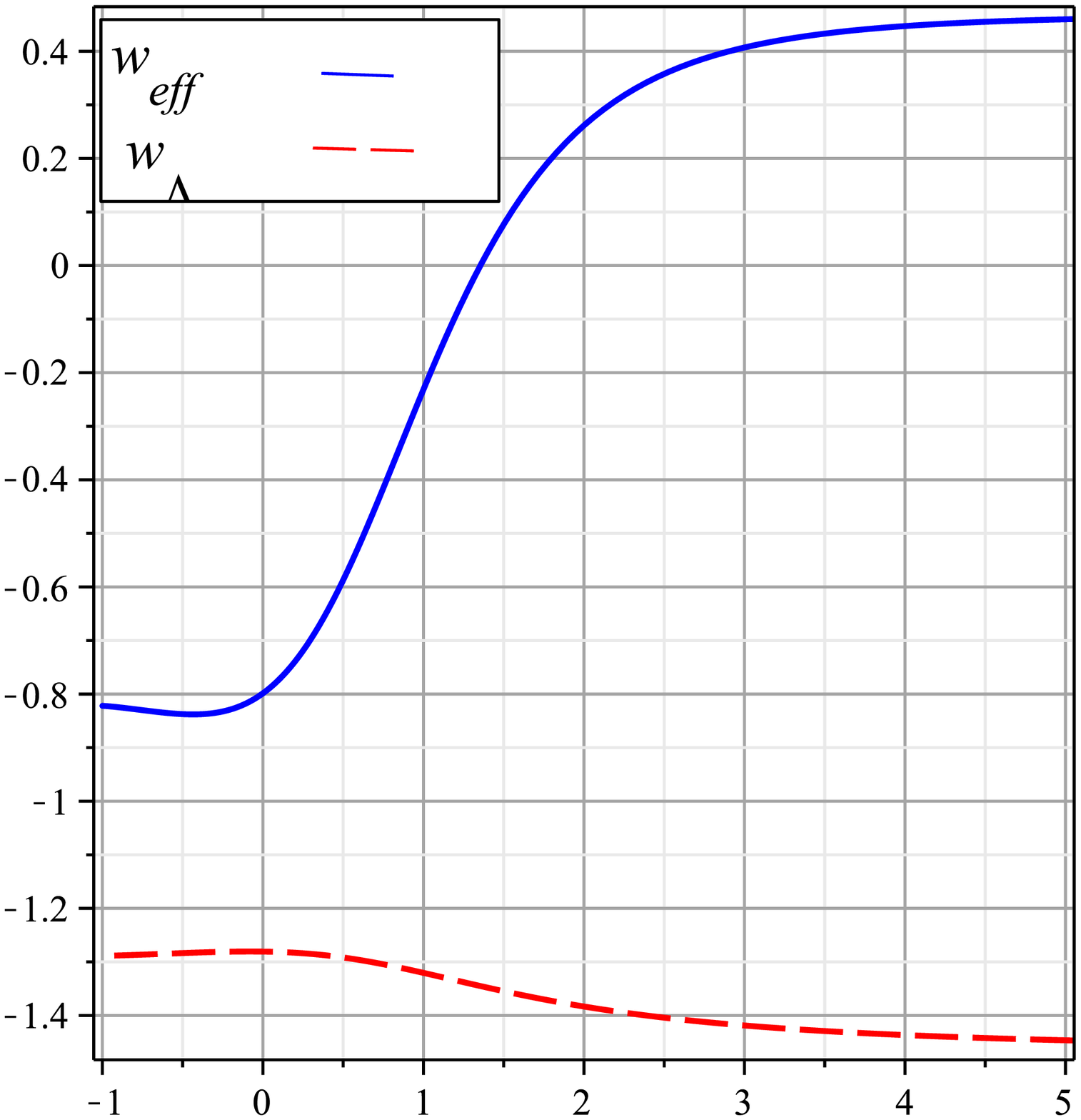}\hspace{0.1 cm}\\
Fig. 4: The graph of effective EoS parameter $w_{eff}$,  as functions of redshift \\
\end{tabular*}\\

Fig 5 shows the trajectories of the statefinder diagrams $\{r,q\}$, $\{s, q\}$ and
$\{r, s\}$. From the graph it can be seen that the best-fitted trajectory passes LCDM state
with ${r, s} = {1, 0}$ sometimes in the past. The current value of the best fitted trajectory
and its location with respect to the LCDM state can also be observed in the ${r, s}$ diagram.\\

\begin{tabular*}{2.5 cm}{cc}
\includegraphics[scale=.3]{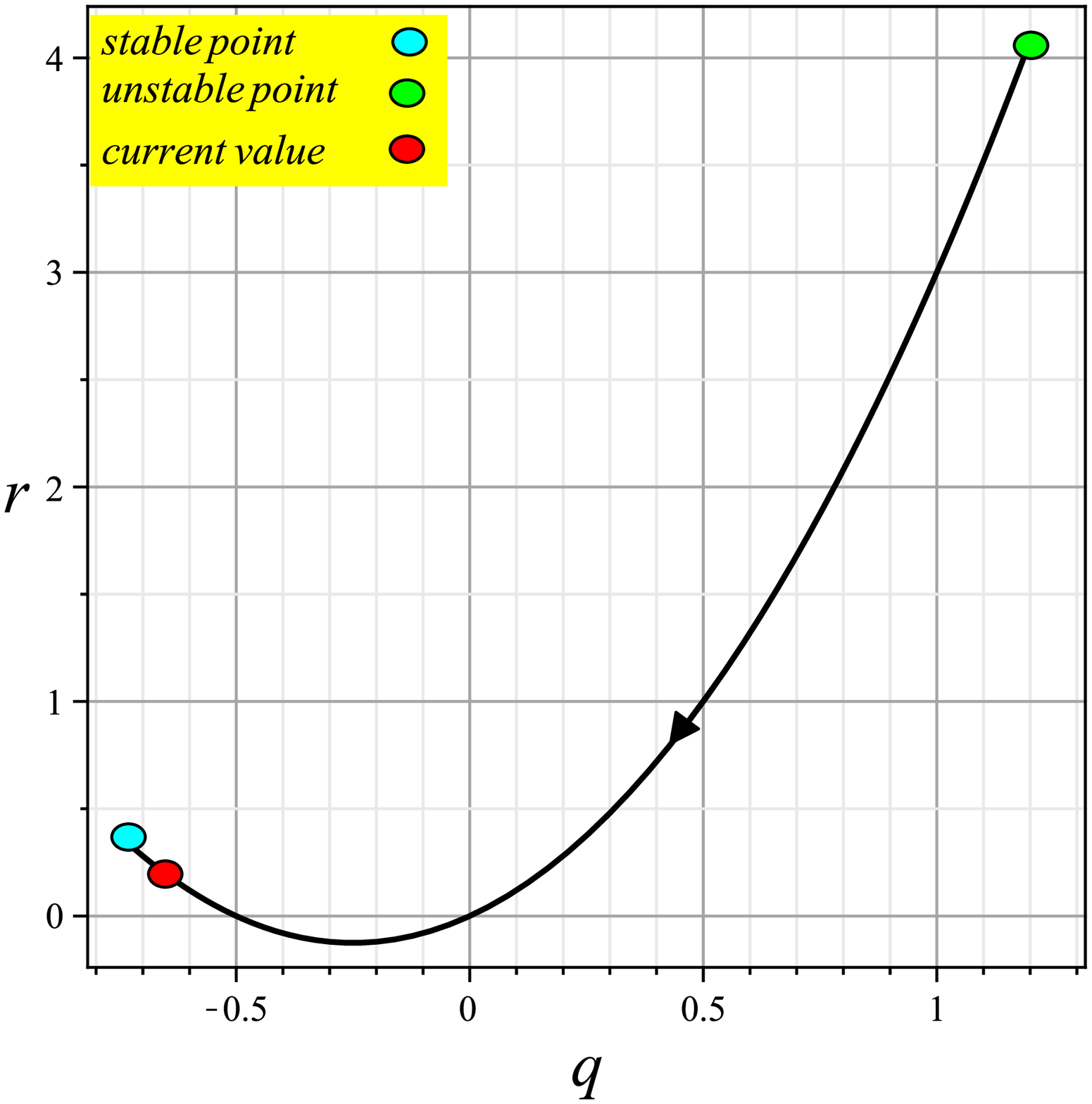}\hspace{0.1 cm} \includegraphics[scale=.3]{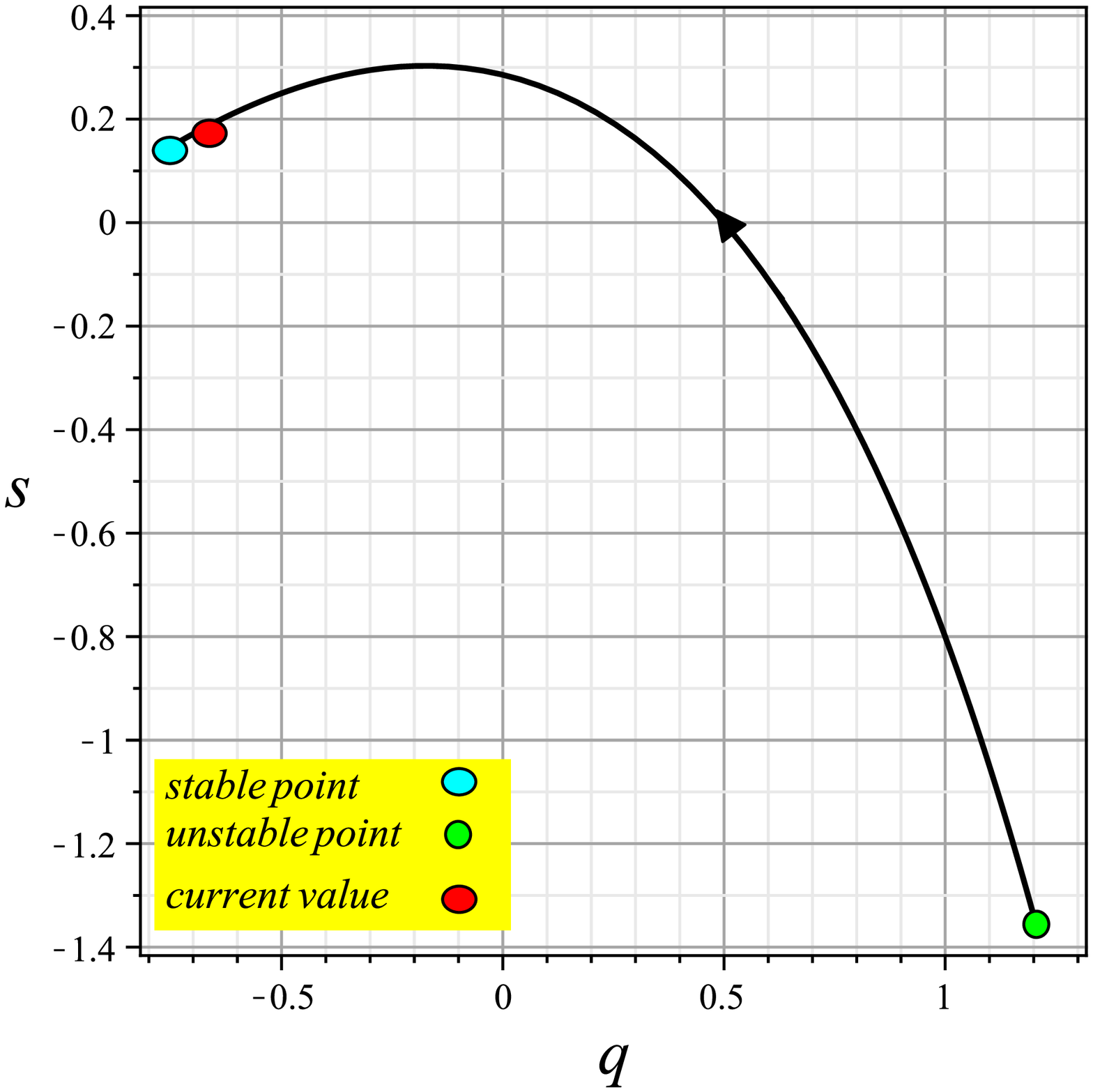}\hspace{0.1 cm} \includegraphics[scale=.3]{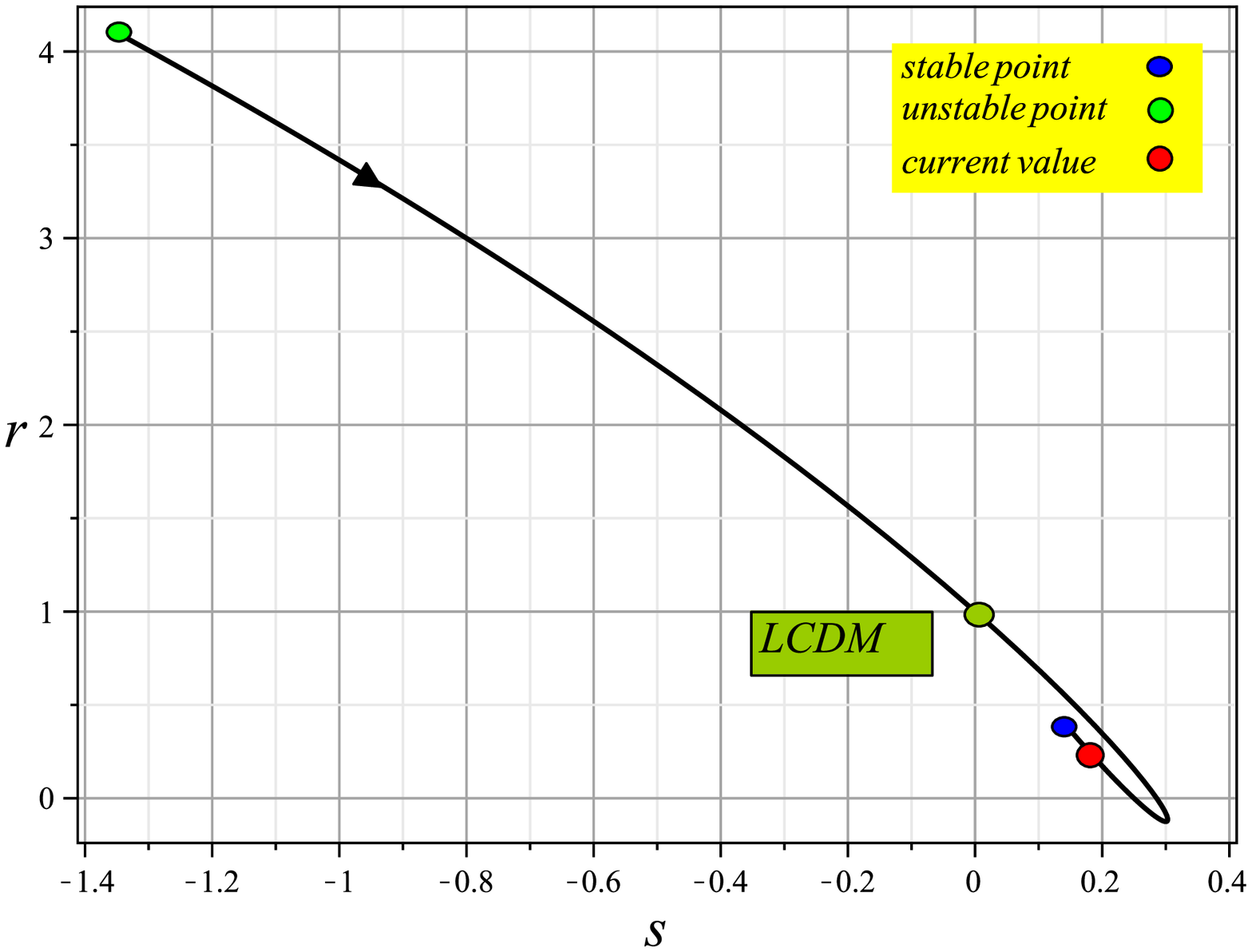}\hspace{0.1 cm} \\
Fig. 5: The plot of statefinder parameters $\{r, q\}$, $\{s, q \}$ and $\{r, s\}$\\
\end{tabular*}\\

Moreover, Fig 6 shows the corresponding dynamical behavior of the satefinder $r$ and $s$ as a function of
$N = ln(a)$.

\begin{tabular*}{2.5 cm}{cc}
\includegraphics[scale=.3]{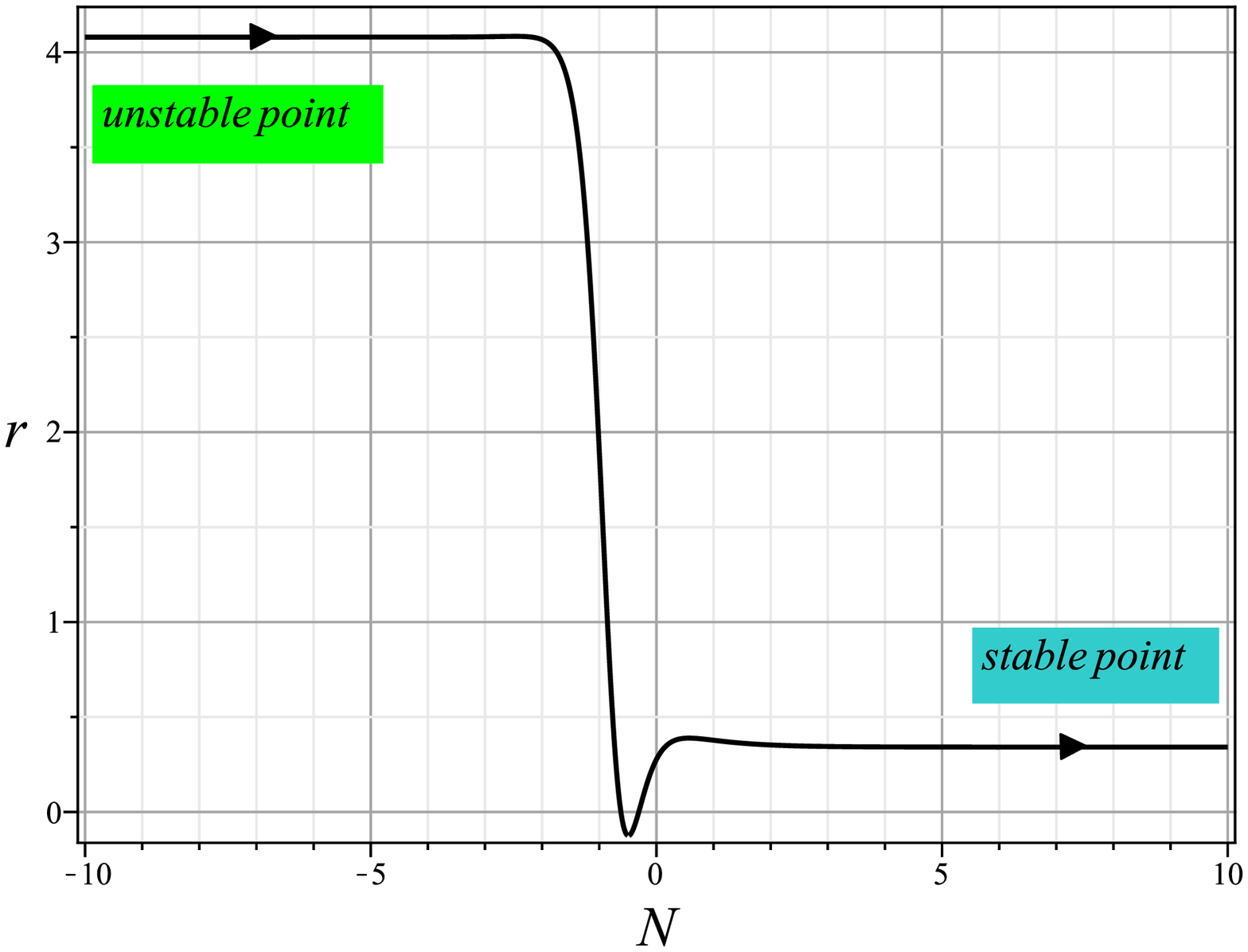}\hspace{0.1 cm}\includegraphics[scale=.3]{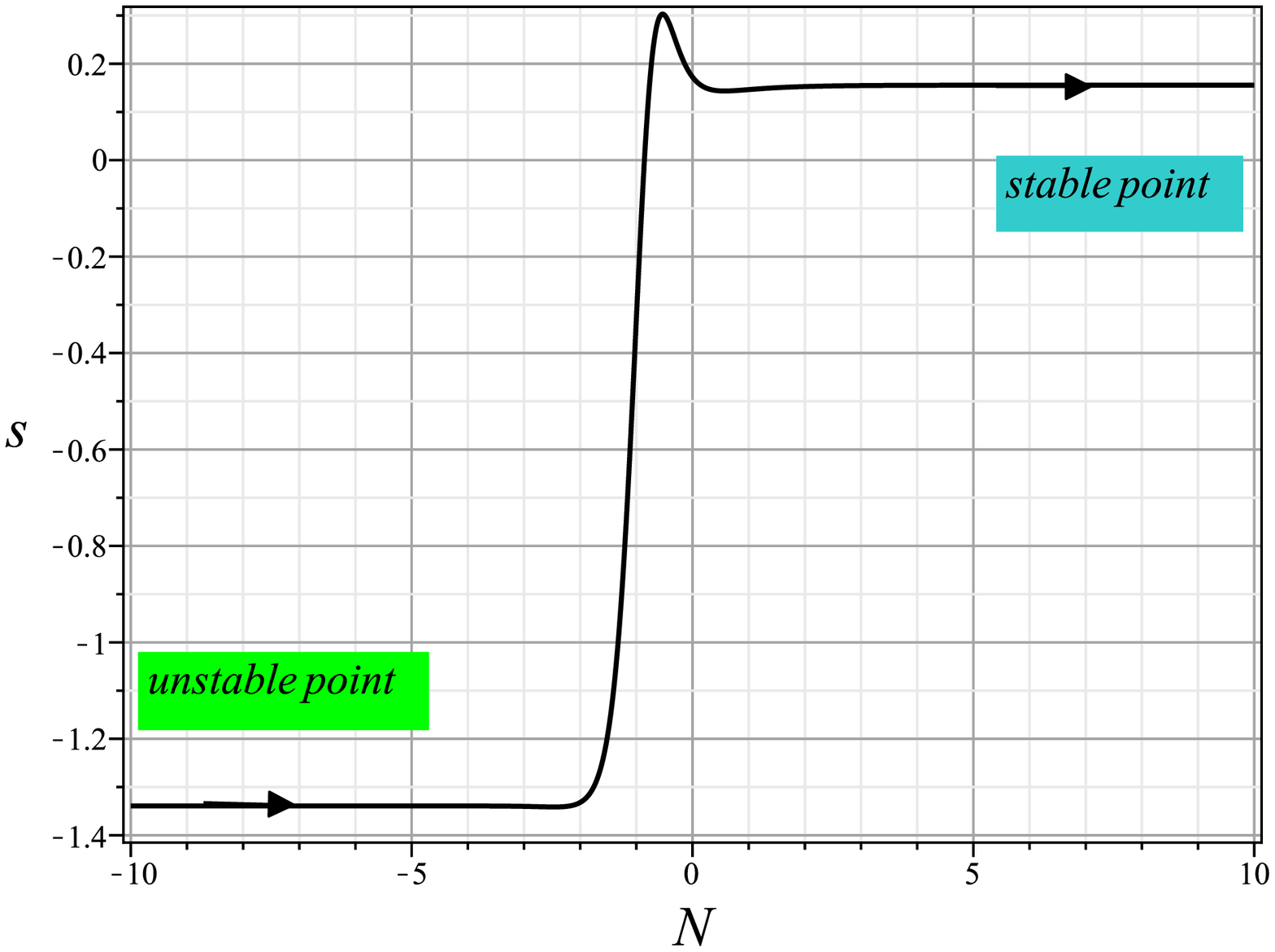}\hspace{0.1 cm}\\
Fig. 6: The plot of statefinder parameters $r$ and $s$  as function of $N=ln(a)$.\\
\end{tabular*}\\
\section{Summary and discussion\label{CONC}}

The paper is designed to investigate the agegraphic dark energy model in Brans-Dicke theory. After the equations have been cast in the appropriate form, from stability analysis and best fitting parameters, one stable and four unstable critical points are obtained. The best fitted trajectory shows that the universe moves from unstable state in the past to the attractor state in future. Testing the model against EoS parameter for agegraphic dark energy implies that universe is always in phantom era. However, if contribution of CDM into the total energy to be taken into account, then the effective EoS parameter reveals that the universe at higher redshifts is in matter dominated epoch and just recently at about $z=0.7$, it enters quintessence phase. The statefinder parameters also show that universe has passed  $LCDM$ state in the past and approaches stable state in future.

\nocite{*}
\bibliographystyle{spr-mp-nameyear-cnd}
\bibliography{biblio-u1}

\end{document}